\documentclass[twocolumn,aps,superscriptaddress]{revtex4}

\usepackage{amssymb}
\usepackage{amsmath}
\usepackage{graphicx}
\usepackage[normalem]{ulem}
\usepackage[dvips]{color}
\usepackage{appendix}
\usepackage{color}

\begin{document}

\title{Bayesian uncertainty quantification for nuclear matter incompressibility}

\author{Jun Xu\footnote{xujun@zjlab.org.cn}}
\affiliation{Shanghai Advanced Research Institute, Chinese Academy of Sciences,
Shanghai 201210, China}
\affiliation{Shanghai Institute of Applied Physics, Chinese Academy
of Sciences, Shanghai 201800, China}
\author{Zhen Zhang\footnote{zhangzh275@mail.sysu.edu.cn}}
\affiliation{Sino-French Institute of Nuclear Engineering and Technology, Sun Yat-Sen University, Zhuhai 519082, China}
\author{Bao-An Li\footnote{Bao-An.Li@tamuc.edu}}
\affiliation{Department of Physics and Astronomy, Texas A$\&$M University-Commerce, Commerce, TX 75429, USA}

\date{\today}

\begin{abstract}
Within a Bayesian statistical framework using the standard Skyrme-Hartree-Fock model, the maximum {\it a posteriori} (MAP) values and uncertainties of nuclear matter incompressibility and isovector interaction parameters are inferred from the experimental data of giant resonances and neutron-skin thicknesses of typical heavy nuclei. With the uncertainties of the isovector interaction parameters constrained by the data of the isovector giant dipole resonance and the neutron-skin thickness, we have obtained $K_0 = 223_{-8}^{+7}$ MeV at 68\% confidence level using the data of the isoscalar giant monopole resonance in $^{208}$Pb measured at the Research Center for Nuclear Physics (RCNP), Japan, and at the Texas A\&M University (TAMU), USA. Although the corresponding $^{120}$Sn data gives a MAP value for $K_0$ about 5 MeV smaller than the $^{208}$Pb data, there are significant overlaps in their posterior probability distribution functions.
\end{abstract}

\maketitle

\section{Introduction}

The incompressibility $K_0$, as a curvature parameter of the nuclear matter equation of state (EOS) at saturation density, is a fundamental quantity for addressing many critical issues in both nuclear physics and astrophysics. It can be measured using multi-messengers from nuclear reaction and structure experiments as well as observations of neutron stars and their mergers. For example, the incompressibility can be constrained by observables sensitive to the EOS at suprasaturation densities, such as the collective flows~\cite{Dan02} or kaon production~\cite{Fuc01} in heavy-ion collisions as well as properties of neutron stars (see, e.g., Refs.~\cite{Lat07,Li20}). It can also be constrained by observables sensitive to the EOS at subsaturation densities, among which is the isoscalar giant monopole resonances (ISGMR), a breathing oscillation mode of a nucleus. Experiments using inelastic scatterings of $\alpha$ particles on nuclei have been carried out at several laboratories to extract the excitation energy of the ISGMR, a sensitive probe of $K_0$. A pioneering work by Blaizot gives the constraint of $K_0 = (210 \pm 30)$ MeV
from analyzing the ISGMR data of $^{40}$Ca, $^{90}$Zr, and $^{208}$Pb~\cite{Bla80}, while early experiments at the TAMU gives $K_0=231 \pm 5$ MeV by comparing the ISGMR in $^{40}$Ca with microscopic calculations using the Gogny interaction~\cite{You99}. Later analyses give a larger range of $K_0 = 220$ MeV to 260 MeV~\cite{Gar18,Col14,Shl06} or around $235 \pm 30$
MeV~\cite{Kha12,Mar18}. Although efforts have been devoted to constraining $K_0$ for four decades~\cite{Bla80,You99,Gar18,Pie10,Sto14,Col14}, its confidence interval has not be accurately determined, mainly due to the uncertainties of the isovector interactions and their correlations with the isoscalar ones~\cite{Col04}. On the other hand, within the same theoretical model, the ISGMR data always favor a smaller $K_0$ value for Sn isotopes than heavy nuclei, leading to the question ``why Sn is so soft" (soft Tin puzzle)~\cite{Pie10,Pie07,Gar07}. Considerable efforts have been devoted to answering this question (see, e.g., Ref.~\cite{Gar18}). For instance, it was proposed that the mutually enhanced magicity effect may play a role in the nuclear matter incompressibility~\cite{Kha09}, but this was later ruled out by the experiments at the RCNP~\cite{Pat13}.

In the present multimessage era of nuclear physics, the uncertainties of the isovector interactions can be much reduced by the experimental data of isovector giant dipole resonances (IVGDR) and neutron-skin thicknesses of heavy nuclei. The IVGDR is an oscillation mode in which neutrons and protons move collectively relative to each other, with its key observables deduced from its strength function as the centroid energy $E_{-1}$ and the electric polarizability $\alpha_D$. Both of them are good probes of the nuclear symmetry energy $E_{sym}$~\cite{Tri08,Rei10,Pie12,Vre12,Roc13b,Col14,Zha14,Roc15,zhangzhen15,zhenghua16,Geb16,Li21} and the isovector nucleon effective mass $m_v^\star$~\cite{Zha16,Kon17,Xu20a}, characterizing the isospin dependence of the nuclear matter EOS and the momentum-dependent single-nucleon potential, respectively. The neutron-skin thickness $\Delta r_{np}$ is the difference in root-mean-square neutron and proton radii, and its values for heavy nuclei have been known as one of the most robust probes of the nuclear symmetry energy at subsaturation densities~\cite{Bro00,Typ01,Chuck01,Fur02,Tod05,Cen09,Zha13,India1,India2,Xu20b,New21}.

All values of the incompressibility $K_0$ mentioned above have been extracted by using the traditional forward-modeling approach and their uncertainties are estimated from the standard $\chi^2$ minimization in fitting the available ISGMR and IVGDR data. It is known that the Bayesian uncertainty quantification has several advantages over the traditional $\chi^2$ fitting in revealing the underlying model parameters \cite{Ken01}. In this work, we perform a Bayesian uncertainty quantification of $K_0$ and isovector nuclear interaction parameters using combined data of ISGMR, IVGDR, and neutron-skin thicknesses of $^{208}$Pb and $^{120}$Sn.  We also quantify the soft Tin puzzle by examining the degree of overlapping of the posterior probability distribution functions (PDFs) of $K_0$ inferred from the ISGMR data of $^{208}$Pb and $^{120}$Sn. We found an incompressibility of $K_0 = 223_{-8}^{+7}$ MeV at 68\% confidence level using the RCNP and TAMU data together. Moreover,  there is a significant overlap between the posterior PDFs of $K_0$ from analyzing the $^{208}$Pb and $^{120}$Sn data, although the MAP value of $K_0$ from the $^{120}$Sn data is about 5 MeV smaller.

\section{Theoretical framework}

We start from the following effective Skyrme interaction between two nucleons at the positions $\vec{r}_1$ and $\vec{r}_2$
\begin{eqnarray}\label{SHFI}
v(\vec{r}_1,\vec{r}_2) &=& t_0(1+x_0P_\sigma)\delta(\vec{r}) \notag \\
&+& \frac{1}{2} t_1(1+x_1P_\sigma)[{\vec{k}'^2}\delta(\vec{r})+\delta(\vec{r})\vec{k}^2] \notag\\
&+&t_2(1+x_2P_\sigma)\vec{k}' \cdot \delta(\vec{r})\vec{k} \notag\\
&+&\frac{1}{6}t_3(1+x_3P_\sigma)\rho^\alpha(\vec{R})\delta(\vec{r}) \notag\\
&+& i W_0(\vec{\sigma}_1+\vec{\sigma_2})[\vec{k}' \times \delta(\vec{r})\vec{k}].
\end{eqnarray}
In the above, $\vec{r}=\vec{r}_1-\vec{r}_2$ and $\vec{R}=(\vec{r}_1+\vec{r}_2)/2$ are the relative and the cental coordinates of the two nucleons, $\vec{k}=(\nabla_1-\nabla_2)/2i$ is the relative momentum operator and $\vec{k}'$ is its complex conjugate acting on the left, and $P_\sigma=(1+\vec{\sigma}_1 \cdot \vec{\sigma}_2)/2$ is the spin exchange operator. The parameters $t_0$, $t_1$, $t_2$, $t_3$, $x_0$, $x_1$, $x_2$, $x_3$, and $\alpha$ can be solved inversely from the macroscopic quantities~\cite{MSL0}, i.e., the saturation density $\rho_0$, the binding energy at the saturation density $E_0$, the incompressibility $K_0$, the isoscalar and isovector nucleon effective mass $m_s^\star$ and $m_v^\star$ at the Fermi momentum in normal nuclear matter, the symmetry energy and its slope parameter at the saturation density $E_{sym}^0$ and $L$, and the isoscalar and isovector density gradient coefficient $G_S$ and $G_V$. The spin-orbit coupling constant is fixed at $W_0=133.3$ MeVfm$^5$. In the present study, the isoscalar nucleon effective mass is fixed as $m_s^\star=0.84m$ with $m$ being the bare nucleon mass, which reproduces both the excitation energies of isoscalar giant quadruple resonance $E_x=10.9 \pm 0.1$ MeV in $^{208}$Pb~\cite{ISGQRex1,ISGQRex2,ISGQRex3,You81,Roc13a} and $E_x=12.7 \pm 0.4$ MeV in $^{120}$Sn~\cite{You81}. With the help of the experimental data of IVGDR, ISGMR, and neutron-skin thickness, we qualify quantitatively the posterior PDFs of $E_{sym}^0$, $L$, $m_v^\star$, and $K_0$ through the Bayesian analysis, while the values of the other macroscopic quantities, which do not affect much the observables discussed here, are kept the same as their empirical ones from the MSL0 interaction~\cite{MSL0}. We note that the $E_{-1}$ and $\alpha_D$ have been shown to be most sensitive to the $E_{sym}^0$, $L$, and $m_v^\star$~\cite{Tri08,Rei10,Pie12,Vre12,Roc13b,Col14,Zha14,Roc15,zhangzhen15,zhenghua16,Geb16,Li21,Zha16,Kon17,Xu20a}, the excitation energy of the ISGMR is most sensitive to the $K_0$~\cite{Bla80,You99,Gar18,Pie10,Sto14,Col14,Kha12,Mar18,Shl06}, and the neutron-skin thickness $\Delta r_{np}$ is most sensitive to the slope parameter of symmetry energy around $\frac{2}{3}\rho_0$ mostly determined by the $E_{sym}^0$ and $L$~\cite{Vin14,X18}.

Based on the Hartree-Fock method, the energy density functional can be obtained from the above Skyrme interaction [Eq.~(\ref{SHFI})]. Here we assume that the nuclei investigated in the present study are spherical and consider only time-even terms in the SHF functional. Using the variational principle, one obtains the single-nucleon Hamlitonian and the Schr\"odinger equation. Solving the Schr\"odinger equation leads to the eigen-energies and wave functions of constituent nucleons, based on which the binding energy, the charge radius, and the neutron-skin thickness can be obtained from this standard procedure~\cite{Vau72}.

The nucleus resonances are studied by applying the random-phase approximation (RPA) method to the Hartree-Fock basis obtained from the standard SHF functional~\cite{Col13}. The operators for the IVGDR and ISGMR are chosen respectively as
\begin{equation}
\hat{F}_{\rm IVGDR} = \frac{N}{A} \sum_{i=1}^Z r_i Y_{\rm 1M}(\hat{r}_i) - \frac{Z}{A} \sum_{i=1}^N r_i Y_{\rm 1M}(\hat{r}_i), \label{QIVGDR}
\end{equation}
and
\begin{equation}
\hat{F}_{\rm ISGMR} = \sum_{i=1}^A r_i^2 Y_{00}(\hat{r}_i),
\end{equation}
where $N$, $Z$, and $A$ are respectively the neutron, proton, and nucleon numbers in a nucleus, $r_i$ is the coordinate of the $i$th nucleon with respect to the center-of-mass of the nucleus, and $Y_{\rm 1M}(\hat{r}_i)$ is the spherical Bessel function with the magnetic quantum number $M$ degenerate in spherical nuclei. Using the RPA method~\cite{Col13}, the strength function
\begin{equation}
S(E) = \sum_\nu |\langle \nu|| \hat{F}  || \tilde{0} \rangle |^2 \delta(E-E_\nu)
\end{equation}
of a nucleus resonance can be obtained, where the square of the reduced matrix element $|\langle \nu|| \hat{F}  || \tilde{0} \rangle |$ represents the transition probability from the ground state $| \tilde{0} \rangle $ to the excited state $| \nu \rangle$. The moments of the strength function can then be calculated from
\begin{equation}
m_k = \int_0^\infty dE E^k S(E).
\end{equation}
The centroid energy $E_{-1}$ of the IVGDR and the electric polarizability $\alpha_D$ can be obtained from the moments of the strength function through the relation
\begin{eqnarray}
E_{-1} &=& \sqrt{m_1/m_{-1}}, \\
\alpha_D &=& \frac{8\pi e^2}{9} m_{-1}.
\end{eqnarray}
For the ISGMR, the RPA results of the excitation energy
\begin{equation}
E_{ISGMR}=m_1/m_0
\end{equation}
are compared with the corresponding experimental data.

The Bayes' theorem states
\begin{equation}
P(M|D) = \frac{P(D|M)P(M)}{\int P(D|M)P(M)dM},
\end{equation}
where $P(M|D)$ is the posterior probability for the model $M$ given the data set $D$, $P(D|M)$ is the likelihood function or the conditional probability for a given theoretical model $M$ to predict correctly the data $D$, and $P(M)$ denotes the prior probability of the model $M$ before being confronted with the data. The denominator of the right-hand side of the above equation is the normalization constant. For the prior PDFs, we choose the model parameters $p_1=E_{sym}^0$ uniformly within $25 \sim 35$ MeV, $p_2=L$ uniformly within $0 \sim 90$ MeV, $p_3=m_v^\star/m$ uniformly within $0.5 \sim 1$, and $p_4=K_0$ uniformly within $200 \sim 300$ MeV. The theoretical results of $d^{th}_1=E_{-1}$, $d^{th}_2=\alpha_D$, $d^{th}_3=\Delta r_{np}$, and $d^{th}_4=E_{ISGMR}$ from the SHF-RPA method are used to calculate the likelihood for these model parameters to reproduce the experimental data $d^{exp}_{1 \sim 4}$ according to
\begin{eqnarray}
&&P[D(d_1,d_2,d_3,d_4)|M(p_1,p_2,p_3,p_4)] \notag\\
&=& \Pi_{i=1}^4 \Bigg \{ \frac{1}{2\pi \sigma_i} \exp\left[-\frac{(d^{th}_i-d^{exp}_i)^2}{2\sigma_i^2}\right] \Bigg\}, \label{llh}
\end{eqnarray}
where $\sigma_{i}$ is the $1\sigma$ error of the data $d_i^{exp}$. The calculation of the posterior PDFs is based on the Markov-Chain Monte Carlo (MCMC) approach using the Metropolis-Hastings algorithm~\cite{Met53,Has70}. Since the MCMC process does not start from an equilibrium distribution, initial samples in the so-called burn-in period have to be thrown away.

\section{Results and discussions}

\begin{widetext}
\begin{table*}\small
  \caption{Experimental data of the centroid energy $E_{-1}$ and electric polarizability $\alpha_D$ in the IVGDR, the neutron-skin thickness $\Delta r_{np}$, the excitation energy $E_{ISGMR}$ in the ISGMR, the average energy per nucleon $E_b$, and the charge radius $R_c$ in $^{120}$Sn and $^{208}$Pb for four data sets used for the Bayesian analysis. For $^{208}$Pb, the $E_{ISGMR}$ data by TAMU and RCNP are used for comparison. Without special notification, the $\Delta r_{np}$ data are deduced from the $L$ values extracted in Ref.~\cite{Xu20b}, while the $\Delta r_{np}$ data for $^{208}$Pb by PREXII is also used in the analysis for comparison.}
    \begin{tabular}{|c|c|c|c|c||c|c|}
   \hline
           & $E_{-1}$ (MeV)& $\alpha_D$ (fm$^3$) & $\Delta r_{np}$ (fm) & $E_{ISGMR}$ (MeV) & $E_b$ (MeV) & $R_c$ (fm) \\
   \hline
   \hline
    $^{208}$Pb-TAMU & $13.46 \pm 0.10$ & $19.6 \pm 0.6$ & $0.170 \pm 0.023$ & $14.17 \pm 0.28$ & $-7.867452 \pm 3\%$ & $5.5010 \pm 3\%$  \\
    $^{208}$Pb-RCNP & $13.46 \pm 0.10$ & $19.6 \pm 0.6$ & $0.170 \pm 0.023$ & $13.9 \pm 0.1$ & $-7.867452\pm 3\%$  & $5.5010 \pm 3\%$ \\
    $^{208}$Pb-RCNP-PREXII & $13.46 \pm 0.10$ & $19.6 \pm 0.6$ & $0.283 \pm 0.071$ & $13.9 \pm 0.1$ & $-7.867452\pm 3\%$  & $5.5010 \pm 3\%$ \\
    $^{120}$Sn & $15.38 \pm 0.10$ & $8.59 \pm 0.37$ & $0.150 \pm 0.017$ & $15.7 \pm 0.1$ & $-8.504548\pm 3\%$  & $4.6543 \pm 3\%$  \\
   \hline
    \end{tabular}
  \label{T1}
\end{table*}
\end{widetext}

\begin{figure*}[ht]
\includegraphics[scale=0.2]{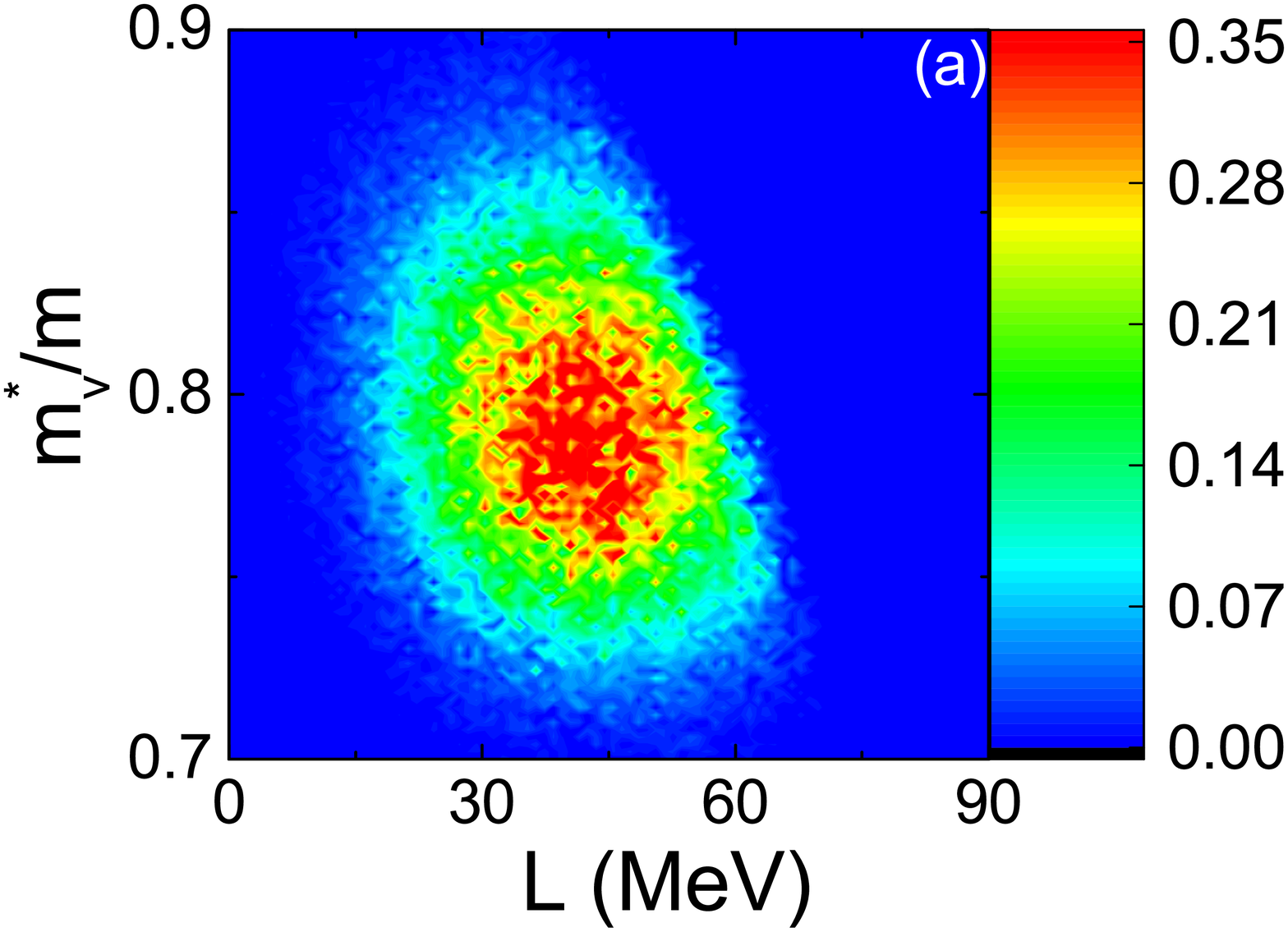}
\includegraphics[scale=0.2]{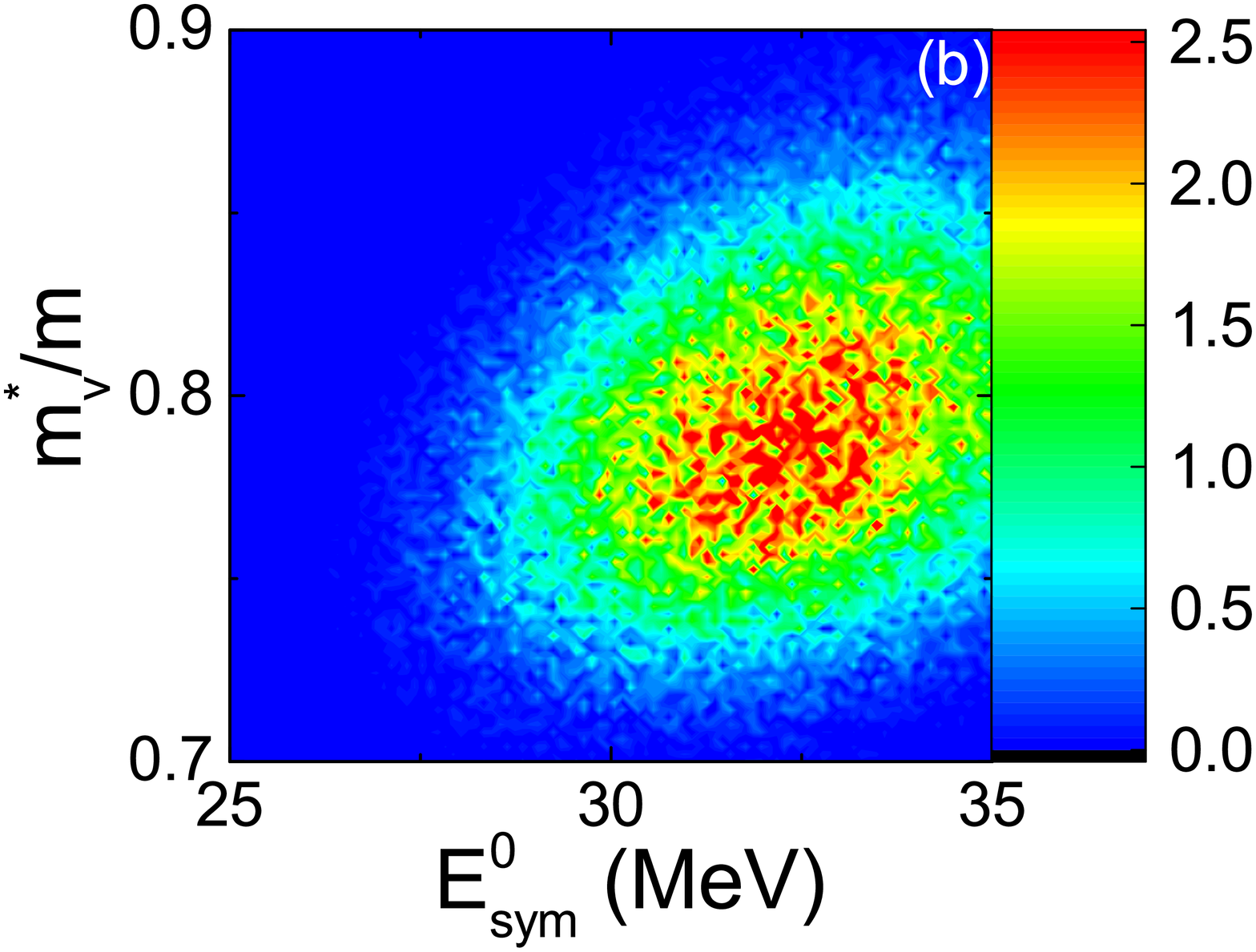}
\includegraphics[scale=0.2]{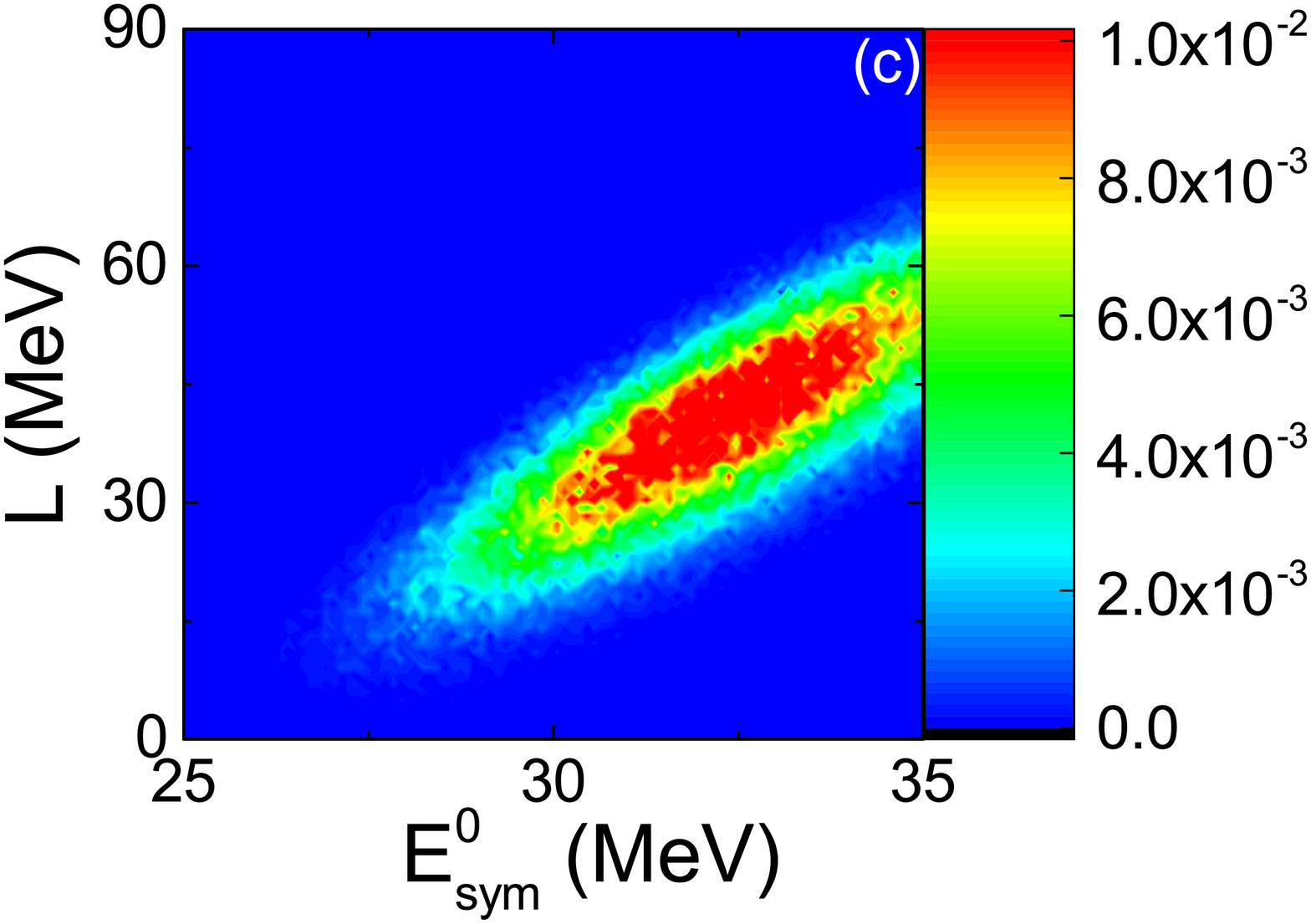}
	\caption{(Color online) Correlated posterior PDFs of isovector interaction parameters from using the $^{208}$Pb-RCNP data in Table \ref{T1} in the $L-m_v^\star/m$ plane (a), the $E_{sym}^0-m_v^\star/m$ plane (b), and the $E_{sym}^0-L$ plane (c), respectively.} \label{fig1}
\end{figure*}

\begin{figure*}[ht]
\includegraphics[scale=0.2]{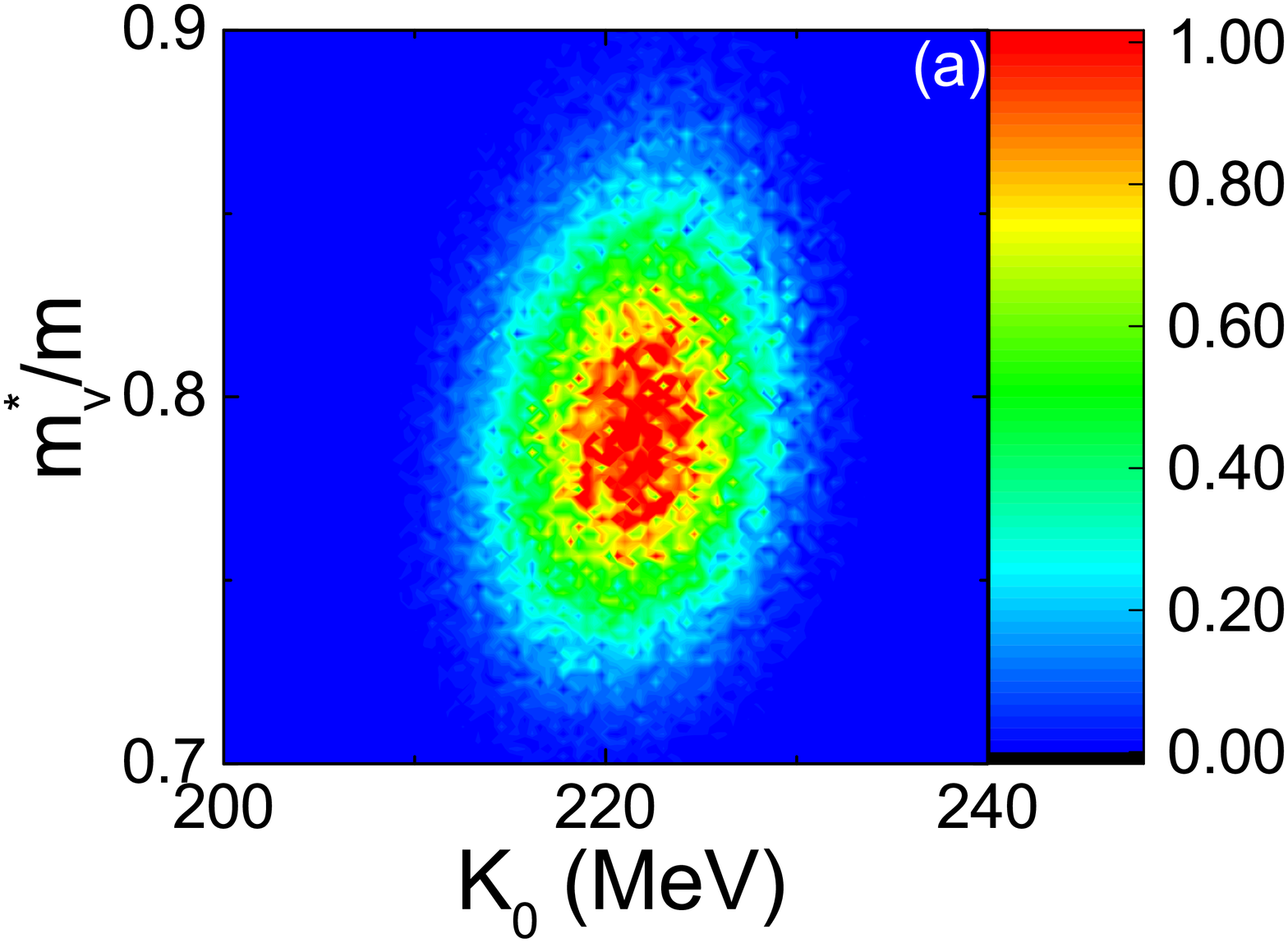}
\includegraphics[scale=0.2]{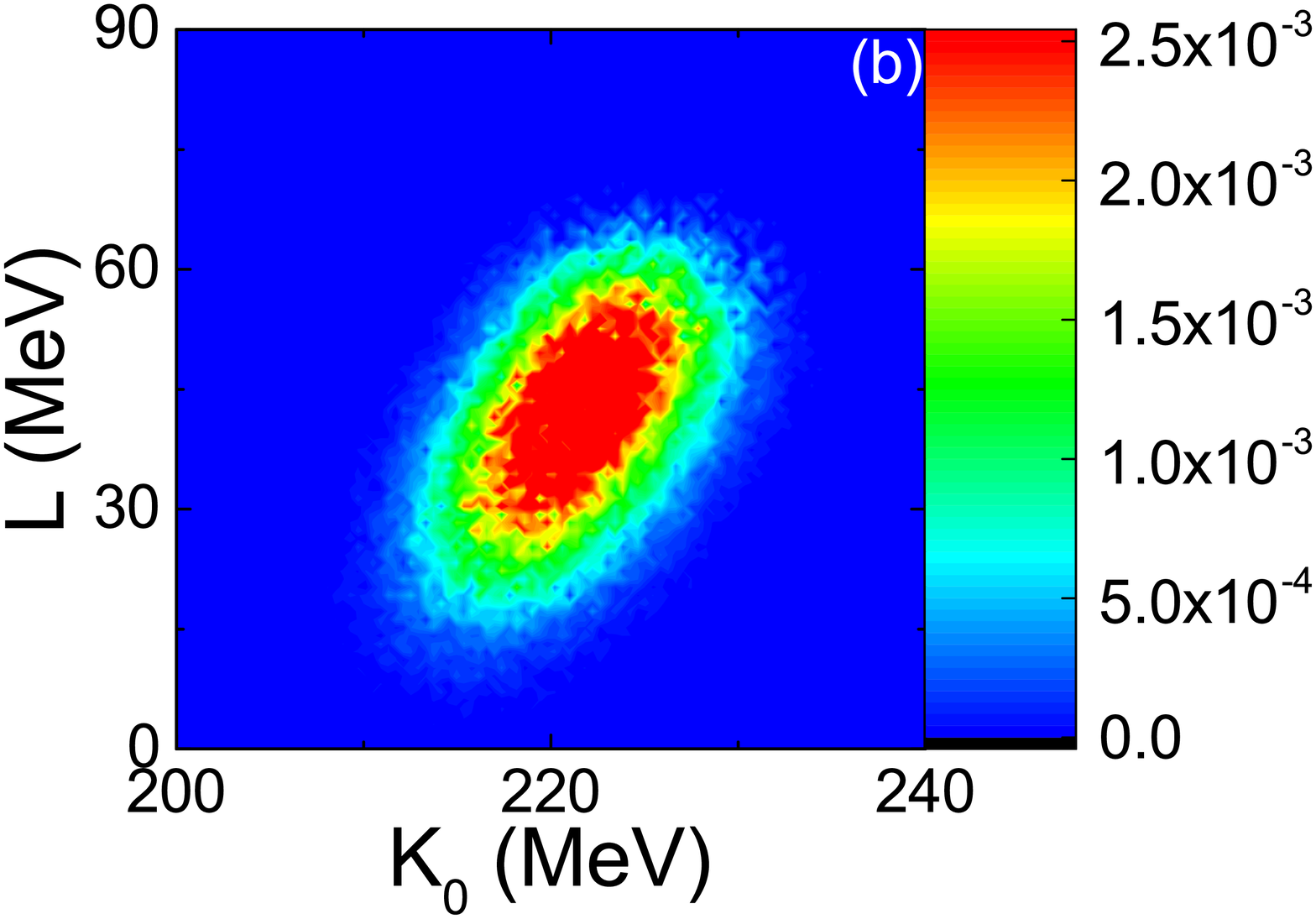}
\includegraphics[scale=0.2]{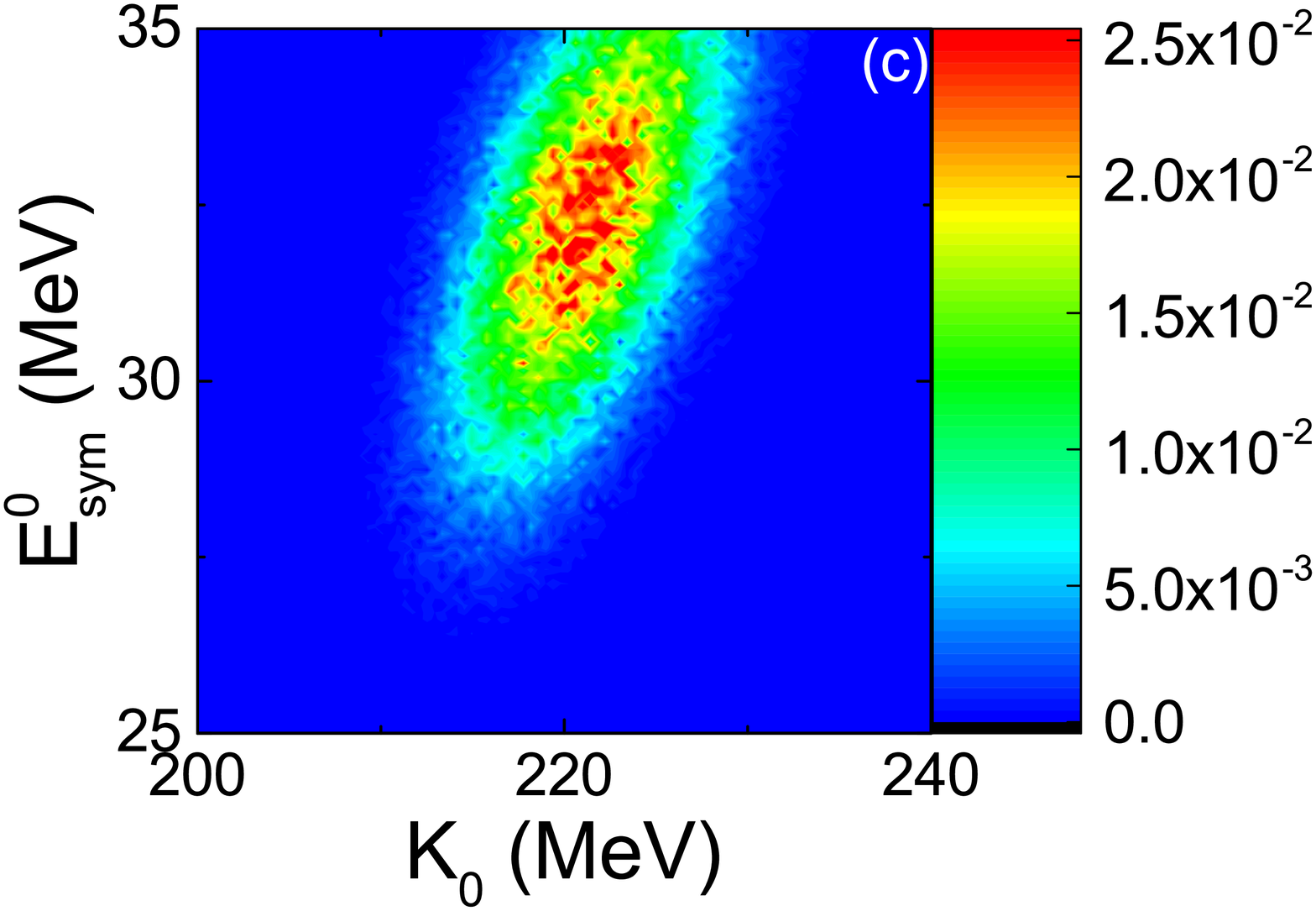}
	\caption{(Color online) Correlated posterior PDFs from using the $^{208}$Pb-RCNP data in Table \ref{T1} in the $K_0-m_v^\star/m$ plane (a), the $K_0-L$ plane (b), and the $K_0-E_{sym}^0$ plane (c), respectively.} \label{fig2}
\end{figure*}

Besides comparing with the experimental data of $E_{-1}$, $\alpha_D$, $\Delta r_{np}$, and $E_{ISGMR}$, we have also used a strong constraint that the theoretical calculation should reproduce the binding energy and charge radius of the corresponding nucleus within $3\%$, an uncertainty range for reasonable SHF parameterization as shown in Ref.~\cite{MSL0}, otherwise the likelihood function [Eq.~(\ref{llh})] is set to 0. This condition guarantees that we are exploring a reasonable space of model parameters. Details of the experimental data for $^{208}$Pb and $^{120}$Sn used in the present study are shown in Table.~I. For $^{208}$Pb, the experimental results of the centroid energy $E_{-1}=13.46$ MeV of the IVGDR from photoneutron scatterings~\cite{IVGDRe}, and the electric polarizability $\alpha_D=19.6 \pm 0.6$ fm$^3$ from polarized proton inelastic scatterings~\cite{Tam11} and with the quasi-deuteron excitation contribution subtracted~\cite{Roc15}, are used in the Bayesian analysis. For $^{120}$Sn, we use the experimental data of $E_{-1}=15.38$ MeV of the IVGDR from photoneutron scatterings~\cite{IVGDRe}, and $\alpha_D=8.59 \pm 0.37$ fm$^3$ from combining the proton inelastic scattering and photoabsorption data~\cite{Has15} and with the quasi-deuteron excitation contribution subtracted~\cite{Roc15}, overlaping with $\alpha_D=8.08 \pm 0.60$ fm$^3$ from the latest data extracted through proton inelastic scatterings~\cite{Bas20a,Bas20b}. The $1\sigma$ error of $E_{-1}$ for both $^{208}$Pb and $^{120}$Sn is chosen to be $0.1$ MeV representing the scale of its uncertainty so far~\cite{IVGDRe}. For the neutron-skin thickness, knowing the uncertainties from various experimental measurements (see, e.g., Ref.~\cite{Xu20b} and references therein), we adopt the predicted values of $\Delta r_{np}=0.170 \pm 0.023$ fm for $^{208}$Pb and $\Delta r_{np}=0.150 \pm 0.017$ fm for $^{120}$Sn from $L(\rho^\star=0.10~{\rm fm}^{-3})=43.7 \pm 5.3$ MeV extracted in Ref.~\cite{Xu20b}, with the latter deduced from the neutron-skin thickness of Sn isotopes from proton elastic scattering experiments~\cite{Ter08}. We also use the latest PREXII data of $\Delta r_{np} = 0.283 \pm 0.071$ fm for $^{208}$Pb from parity violating electron-nucleus scatterings~\cite{PREXII} in a different data set. For the excitation energy of the ISGMR from inelastic scatterings of $\alpha$ particles, we use $E_{ISGMR}=15.7 \pm 0.1$ MeV for $^{120}$Sn by the RCNP, Osaka University~\cite{SnRCNP}, and for $^{208}$Pb we use both $E_{ISGMR}=14.17 \pm 0.28$ MeV by the TAMU~\cite{You99} and $E_{ISGMR}=13.9 \pm 0.1$ MeV by the RCNP~\cite{PbRCNP} for comparison. The experimental data of the binding energies and charge radii of $^{208}$Pb and $^{120}$Sn are taken from Refs.~\cite{Aud03,Ang04}.

Shown in Fig.~\ref{fig1} are the correlated posterior PDFs of isovector interaction parameters $m_v^\star/m$, $L$, and $E_{sym}^0$ using the $^{208}$Pb-RCNP data set in Table \ref{T1}. There is no strong correlation between $m_v^\star/m$ and $L$ or $E_{sym}^0$, while a strong positive correlation is observed between $L$ and $E_{sym}^0$. The latter is due to the constraint from the IVGDR data, as discussed in Refs.~\cite{Xu20a,Xu20b,Xu21}. The correlated PDFs with only IVGDR and neutron-skin thickness can be found in Fig.~3 of Ref.~\cite{Xu21}. It is seen that incorporating the constraint from the ISGMR data does not affect much the correlated PDFs of the isovector interaction parameters.

Correlations between the incompressibility $K_0$ and the isovector interaction parameters from using the $^{208}$Pb-RCNP data are displayed in Fig.~\ref{fig2}. It is seen that the correlations between the $K_0$ and the isovector parameters are generally weaker compared with the strong $L-E_{sym}^0$ correlation shown in Fig.~\ref{fig1} (c). Nevertheless, the positive correlation between $K_0$ and $L$ or $E_{sym}^0$ could lead to uncertainties in constraining $K_0$ (see, e.g., discussions in Ref.~\cite{Col14}). This actually leads to some correlation between $K_0$ and the curvature parameter $K_{sym}$ of the symmetry energy as well, though $K_{sym}$ is not an independent variable in the present framework but can be calculated from other quantities, e.g., $L$, $E_{sym}^0$, etc. All correlated posterior PDFs related to $E_{sym}^0$ (Fig.~\ref{fig1}(b), Fig.~\ref{fig1}(c), and Fig.~\ref{fig2}(c)) hit its upper boundary, limited by the empirical prior range of $E_{sym}^0$ deduced from various earlier analyses~\cite{Li13,Oer17}.

\begin{figure*}[ht]
\includegraphics[scale=0.4]{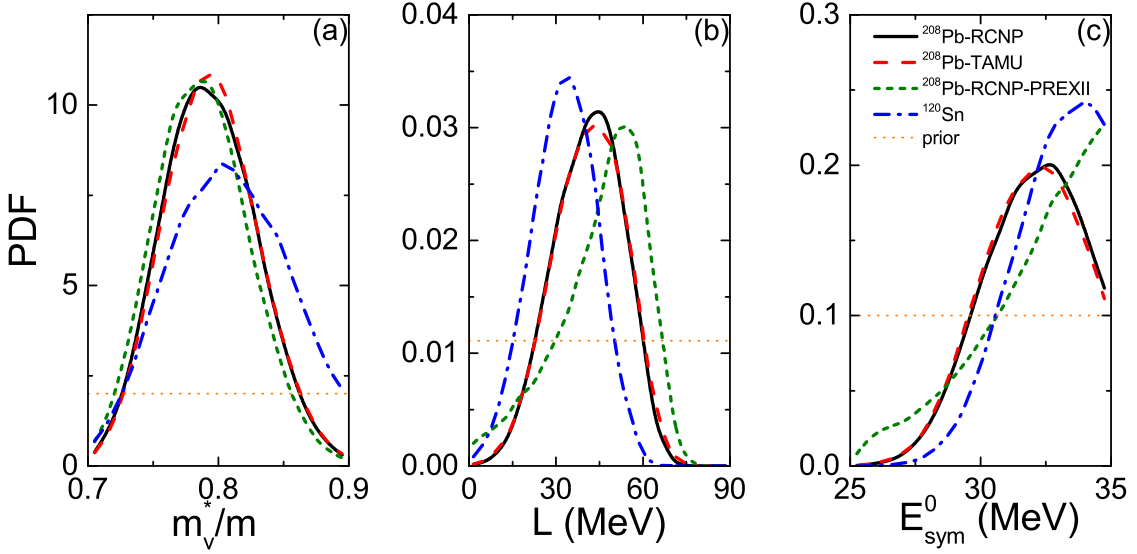}
	\caption{(Color online) Prior and posterior PDFs of $m_v^\star/m$ (a), $L$ (b), and $E_{sym}^0$ (c) from different data sets in Table \ref{T1}.} \label{fig3}
\end{figure*}

\begin{figure*}[ht]
\includegraphics[scale=0.4]{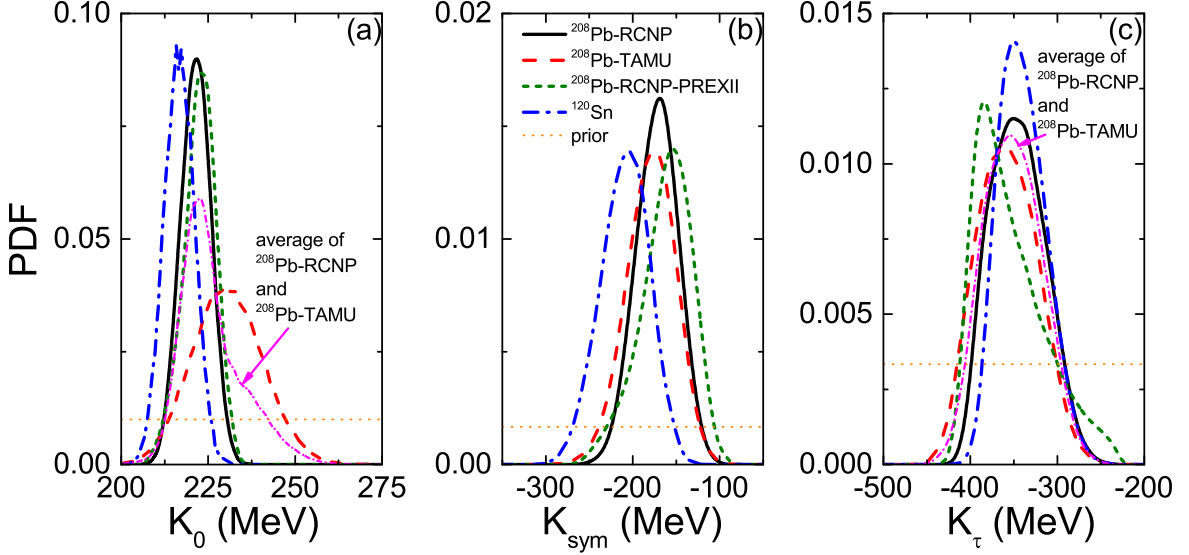}
	\caption{(Color online) Prior and posterior PDFs of $K_0$ (a), $K_{sym}$ (b), and $K_\tau$ (c) from different data sets in Table \ref{T1}.} \label{fig4}
\end{figure*}

The posterior PDFs of isovector interaction parameters from using different data sets are compared in Fig.~\ref{fig3}, where their prior distributions are also shown by dotted lines. The difference between the $^{208}$Pb-RCNP and $^{208}$Pb-TAMU data sets is in the ISGMR data, and the resulting PDFs of isovector interaction parameters are thus almost the same. Incorporating the PREXII data of neutron-skin thickness to the $^{208}$Pb-RCNP data leads to a larger $L$ and $E_{sym}^0$, and the MAP value of $L$ changes from about 44 MeV to about 54 MeV. Although the PREXII neutron-skin data itself leads to a large $L$ (see, e.g., Ref.~\cite{Ree21}), its effect in the combined data analysis is small as the IVGDR data with much smaller errors bar have a stronger constraining power at subsaturation densities. Combining the neutron-skin and IVGDR data, it is seen that the $^{120}$Sn data lead to a slightly larger $m_v^\star/m$, a smaller $L$, and also a different $E_{sym}^0$. Interestingly,  there are significant overlaps between the PDFs from analyzing the $^{120}$Sn and $^{208}$Pb data.

The PDFs of $K_0$ from using the four data sets are shown in Fig.~\ref{fig4} (a), where its uniform prior distribution is also displayed. At 68\% confidence level, the $^{208}$Pb-RCNP data gives $K_0=221 \pm 4$ MeV, while the $^{208}$Pb-TAMU data gives $K_0=230 \pm 10$ MeV, since a larger $E_{ISGMR}$ with a larger error bar is used in the latter case as shown in Table \ref{T1}. Assuming the data from RCNP and TAMU are equally reliable, the average incompressibility from the two analyses is $K_0 = 223_{-8}^{+7}$ MeV. As shown in Fig.~\ref{fig4} (b) and (c), the prior $K_{sym}$ and $K_\tau$ are uniformly distributed within $[-500, 100]$ MeV and $[-500, -200]$ MeV, respectively, with~\cite{Che09}
\begin{equation}
K_\tau = K_{sym}-6L-\frac{J_0}{K_0}L
\end{equation}
reflecting the isospin dependence of the isobaric incompressibility of asymmetric nuclear matter along its saturation line, and $J_0$ being the skewness of symmetric nuclear matter EOS. The ranges of $K_{sym}$ and $K_\tau$ from the uncertainties of isovector interactions are narrowed down by comparing with the data in Table~\ref{T1}, helping to put a more stringent constraint on $K_0$. Compared with the $^{208}$Pb-RCNP data, the $^{208}$Pb-RCNP-PREXII data set gives a larger $L$ as shown in Fig.~\ref{fig3} (b), and also a larger $K_{sym}$ as shown in Fig.~\ref{fig4} (b), leading to a smaller $K_\tau$ as shown in Fig.~\ref{fig4} (c).

Indeed, the $^{120}$Sn data gives a MAP value of $K_0$ about 5 MeV smaller than the $^{208}$Pb data, qualitatively consistent with the soft Tin puzzle. Interestingly, however,
quantitatively there is a significant overlap in the PDFs of $K_0$ from analyzing the $^{120}$Sn and $^{208}$Pb data.
In addition, the $^{120}$Sn data give a smaller $K_{sym}$ but a similar $K_\tau$, compared with the $^{208}$Pb data. At 68\% confidence level, the $^{208}$Pb-RCNP data gives $K_\tau = -350^{+35}_{-25}$ MeV, and the $^{208}$Pb-TAMU data gives $K_\tau = -360 \pm 35$ MeV. Their average is $K_\tau = -355 \pm 30$ MeV.\\

\section{Conclusions}

We performed a Bayesian uncertainty quantification for the incompressibility $K_0$ of nuclear matter and the three isovector interaction parameters using the experimental data of isoscalar giant monopole resonances, isovector dipole resonances, and the neutron-skin thickness mainly from RCNP and TAMU. Taking the average results extracted from analyzing the RCNP and TAMU data for $^{208}$Pb, the isoscalar and isovector parts of the nuclear incompressibility are constrained to $K_0 = 223_{-8}^{+7}$ MeV and $K_\tau = -355 \pm 30$ MeV, respectively, at 68\% confidence level. We also quantified the soft Tin puzzle. Although the resulting $K_0$ is about 5 MeV smaller from analyzing the $^{120}$Sn data than the $^{208}$Pb data, there is a significant overlap in their posterior PDFs.

\begin{acknowledgments}

We thank Umesh Garg for many communications and discussions on the physics of nuclear giant resonances, and Wen-Jie Xie for helpful discussions on the Bayesian analysis. JX acknowledges the National Natural Science Foundation of China under Grant No. 11922514. ZZ acknowledges the National Natural Science Foundation of China under Grant No. 11905302. BAL acknowledges the U.S. Department of Energy, Office of Science, under Grant No. DE-SC0013702, and the CUSTIPEN (China-U.S. Theory Institute for Physics with Exotic Nuclei) under the U.S. Department of Energy Grant No. DE-SC0009971.

\end{acknowledgments}

\end{document}